\documentclass[aps,showpacs,two column,showkeys,superscriptaddress,preprintnumbers]{revtex4}
\usepackage{epsfig,amssymb,subfigure,bm,dsfont}
\usepackage{chngcntr}
\usepackage{amsmath}
\usepackage{gensymb}
\usepackage{txfonts}
\usepackage[titletoc]{appendix}
\usepackage{amssymb}
\usepackage{tabularray}
\usepackage{array}
\usepackage{mathrsfs}
\usepackage{subfigure,overpic}
\usepackage{dcolumn}
\usepackage{graphicx}
\usepackage{epstopdf}
\usepackage{float}
\usepackage{balance}
\usepackage{setspace} 
\usepackage{lipsum}
\usepackage[colorlinks,
            linkcolor=blue,
            anchorcolor=blue,
            citecolor=blue]{hyperref}
\usepackage{setspace}
\usepackage{multirow}
\usepackage{makecell}
\usepackage[normalem]{ulem} % 波浪线包
\usepackage{xcolor}
\usepackage{pifont}
\usepackage{graphicx}
\usepackage{tikz}     % 更高级的控制
\usepackage[capitalize]{cleveref}
\usetikzlibrary{decorations.pathmorphing}
\definecolor{myblue}{rgb}{0,0,1}
\definecolor{myred}{rgb}{1,0,0}
\definecolor{myblack}{rgb}{0,0,0}
\definecolor{mymagenta}{rgb}{1,0,1}
\newcommand{\safebigding}[2][1.5]{  \ensuremath{\vcenter{\hbox{\scalebox{#1}{\ding{#2}}}}}}

\begin{document}

\title{N\'{e}el vector and Rashba SOC effects on RKKY interaction in 2D $d$-wave altermagnets}

\author{Hou-Jian Duan}
\email{dhjphd@163.com}
\affiliation{Guangdong Basic Research Center of Excellence for Structure and Fundamental Interactions of Matter, Guangdong Provincial Key Laboratory of Quantum Engineering and Quantum Materials, School of Physics, South China Normal University, Guangzhou 510006, China}
\affiliation{Guangdong-Hong Kong Joint Laboratory of Quantum Matter, Frontier Research Institute for Physics, South China Normal University, Guangzhou 510006, China}
\author{Miao-Sheng Fang}
\affiliation{China Mobile Communications Group Guangdong Co., Ltd. Dongguan Branch, Dongguan 523808, China}
\author{Ming-Xun Deng}
\affiliation{Guangdong Basic Research Center of Excellence for Structure and Fundamental Interactions of Matter, Guangdong Provincial Key Laboratory of Quantum Engineering and Quantum Materials, School of Physics, South China Normal University, Guangzhou 510006, China}
\affiliation{Guangdong-Hong Kong Joint Laboratory of Quantum Matter, Frontier Research Institute for Physics, South China Normal University, Guangzhou 510006, China}
\author{Ruiqiang Wang}
\email{wangruiqiang@m.scnu.edu.cn}
\affiliation{Guangdong Basic Research Center of Excellence for Structure and Fundamental Interactions of Matter, Guangdong Provincial Key Laboratory of Quantum Engineering and Quantum Materials, School of Physics, South China Normal University, Guangzhou 510006, China}
\affiliation{Guangdong-Hong Kong Joint Laboratory of Quantum Matter, Frontier Research Institute for Physics, South China Normal University, Guangzhou 510006, China}
\author{Mou Yang}
\affiliation{Guangdong Basic Research Center of Excellence for Structure and Fundamental Interactions of Matter, Guangdong Provincial Key Laboratory of Quantum Engineering and Quantum Materials, School of Physics, South China Normal University, Guangzhou 510006, China}
\affiliation{Guangdong-Hong Kong Joint Laboratory of Quantum Matter, Frontier Research Institute for Physics, South China Normal University, Guangzhou 510006, China}

\begin{abstract}
Altermagnets possess two key features: non-relativistic alternating spin splitting (i.e., altermagnetism) and a material-dependent N\'{e}el vector. The former naturally coexists with Rashba spin-orbit coupling (SOC) in real materials on substrates, prompting the question of how SOC affects the magnetic properties of altermagnets. The latter is crucial for information storage, making it essential to determine its orientation. To address these issues, we study the Ruderman-Kittel-Kasuya-Yosida (RKKY) interaction in two-dimensional (2D) $d$-wave altermagnets by independently varying the N\'{e}el vector orientation and the SOC strength. Our results demonstrate that the N\'{e}el vector orientation can be accurately determined via the Ising term without SOC, or qualitatively inferred via the Dzyaloshinskii-Moriya (DM) terms with SOC. Moreover, we observe a novel DM component distinct from previous reports, whose emergence is attributed to the synergy between altermagnetism and SOC. Additionally, through tuning SOC strength, we reveal the evolution of the RKKY spin models governed by five distinct mechanisms: the spin model may be determined solely by altermagnetism, solely by SOC, or solely by the kinetic term; alternatively, altermagnetism may coincidentally yield the same moderately anisotropic spin model as SOC, or compete with SOC to produce a spin model with maximal anisotropy. Beyond SOC strength, which mechanism operates also relies on the N\'{e}el vector orientation and impurity configurations. All results are numerically verified. These findings---which were inaccessible in prior studies due to the limitations of first-order SOC expansion and fixed N\'{e}el vector orientation---provide important new insights into the magnetic properties of altermagnets.
\end{abstract}

\maketitle

%%%%%%%%%%%%%%%%%%%%%%%%%%%%%%%%%%%%%
\section{introduction}
Recently, a new class of magnetic materials termed altermagnets was discovered based on non-relativistic spin symmetry groups \cite{altermagnet1,altermagnet2}. Altermagnets unify characteristics of antiferromagnets and ferromagnets: they preserve a vanishing net magnetization while exhibiting strong $\mathcal{T}$-symmetry breaking and non-relativistic alternating spin splitting (i.e., altermagnetism) in their band structure. This spin-splitting mechanism differs fundamentally from conventional relativistic SOC, enabling novel forms of spin-momentum locking.
\par
So far, numerous altermagnetic candidates have been predicted or confirmed. The family of 3D materials is extensive, including ${\rm MnTe}$, ${\rm Mn_5Si_3}$, ${\rm RuO_2}$ and others \cite{altermagnet3,MnTe1,MnTe2,MnTe3,CrSb0,CrSb1,CrSb2,CrSb3,Mn5Si3_1,RuO2_1,RuO2_2,RuO2_3,RuO2_4,RuO2_5}. In contrast, viable 2D altermagnets remain scarce, with representative examples such as ${\rm K}$-doped ${\rm V_2Se_2O}$ \cite{altermagnet4}, ${\rm Rb}$-doped ${\rm V_2Te_2O}$ \cite{altermagnet5}, non-van der Waals ${\rm FeX}$ (${\rm X = S}$, ${\rm Se}$) semiconductors \cite{altermagnet6}, monolayer ${\rm MnTeMoO_6}$ and ${\rm P_2H_8(NO_4)_2}$ \cite{altermagnet7}. Notably, most predicted 2D altermagnets exhibit $d$-wave symmetry. These materials have stimulated extensive research into their electrical and optical properties \cite{altermagnet3, altermagnet11}, revealing phenomena such as non-relativistic charge-spin conversion \cite{phenomena1,phenomena2,phenomena3,phenomena4,phenomena5,phenomena6,RuO2_1,RuO2_3}, anomalous Hall and Nernst effects \cite{MnTe1,RuO2_2,RuO2_5,hall1,hall2,hall3,hall4,hall5}, giant magnetoresistance \cite{resis1,resis2,resis3,resis4,resis5}, large piezomagnetism \cite{strain1,strain2}, and magneto-optical Kerr effects \cite{opt1,opt2,opt3}. Such studies establish altermagnets as promising candidates for spintronics, data storage, and energy-efficient electronics.
\par
In contrast, research into their magnetic properties---especially the RKKY interaction---has been much less extensive. Indeed, only a handful of studies have addressed the RKKY interaction in altermagnets \cite{alterRKKY1, alterRKKY2, alterRKKY3}, while prior RKKY research was predominantly focused on relativistic systems (probing intrinsic properties and topological states \cite{chang1,chang2,surface1,surface2,surface3,surface4,NLSM,weylpoints1,weylpoints2,weylpoints3,weylpoints4,chiralmodel,tilt1,tilt2,semi,optRKKY1,optRKKY2,optRKKY3}). Although these few works on altermagnets have identified key features such as the direction-dependent beating pattern of the RKKY interaction from the $d$-wave spin splitting \cite{alterRKKY2} and the selective, anisotropic control of the DM component via circularly polarized light \cite{alterRKKY3}, they exhibit important limitations. For instance, Lee's work \cite{alterRKKY1} neglects SOC entirely, resulting in purely Heisenberg and Ising terms. Meanwhile, Refs. \cite{alterRKKY2,alterRKKY3} include Rashba SOC but treat it only to first order in their derivations, thereby overlooking its full impact on the collinear RKKY components and resulting in DM terms that are confined to the in-plane directions. Moreover, all these existing studies assume a fixed N\'{e}el vector orientation (typically out-of-plane). This assumption not only restricts the resulting RKKY spin model to a single, fixed form but also precludes any investigation into the influence of the N\'{e}el vector on the DM term, particularly the possibility of an out-of-plane DM component. These shortcomings highlight the need for a thorough revisit of the RKKY interaction in $d$-wave altermagnets.
\par
To address these shortcomings, our work is driven by three main goals: (1) To go beyond the first-order SOC treatment and reveal the complete influence of Rashba SOC---including higher-order contributions---on the RKKY interaction through full numerical calculations. (2) To provide a platform for investigating the interplay between non-relativistic (altermagnetism) and relativistic (Rashba SOC) effects, which is expected to generate novel RKKY components and enable richer spin textures (such as diverse RKKY spin models and novel DM components). (3) To establish a clear connection between the N\'{e}el vector orientation and the RKKY interaction. In the long-range limit (where the low-energy band structure is unaffected by impurities \cite{longrange}), this connection would allow the RKKY interaction to serve as a magnetic probe for determining the N\'{e}el vector orientation, complementing existing electrical methods \cite{neelprobe}.
\par
The paper is organized as follows. Section II introduces the minimal model, presents the Fermi surface and spin texture, and outlines the method for computing the RKKY interaction. In Section III, we investigate the influence of the N\'{e}el vector on the RKKY interaction both without and with SOC, uncover the synergistic effects arising from the coexistence of altermagnetism and SOC, and track the evolution of the RKKY spin model with SOC strength. We conclude with a summary of our findings.

%%%%%%%%%%%%%%%%%%%%%%%%%%%%%%%%%%%%%%%%%%%%%%%%%%%%%%%%%
\section{Model and Method}
\begin{figure}[!htb]
\centering \includegraphics[width=0.5\textwidth]{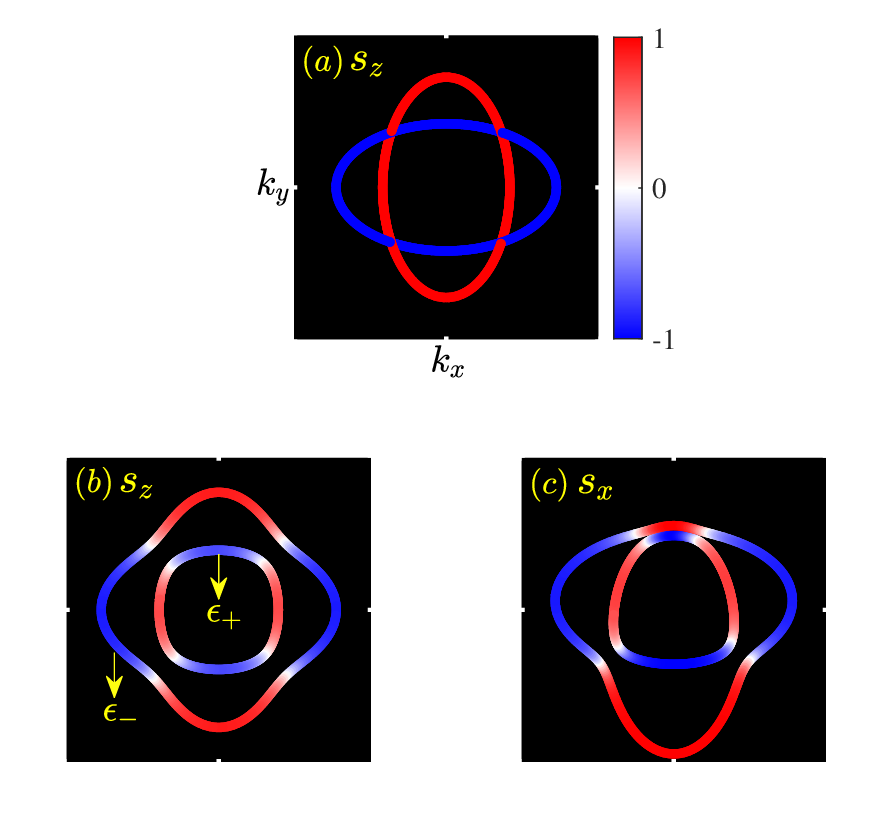}
\caption{(Color online) The Fermi surfaces and the spin expectation $s_i$ (whose value is evaluated by color) in altermagnet for different system parameters: (a) $\alpha=0$, $\mathbf{n}=(0,0,1)$; (b) $\alpha=0.2$ ${\rm eV}$ ${\rm \AA}$, $\mathbf{n}=(0,0,1)$. (c) $\alpha=0.2$ ${\rm eV}$ ${\rm \AA}$, $\mathbf{n}=(1,0,0)$. Other parameters are set as: $\beta=0.5$, $D=5.44$ ${\rm eV}$ ${\rm \AA^2}$, $u_F=0.05$ ${\rm eV}$. }
\end{figure}
We adopt a model of 2D $d$-wave altermagnets similar to Ref. \cite{subs}, the corresponding low-energy Hamiltonian is given by
\begin{eqnarray}\label{eq1}
H=Dk^{2}\sigma _{0}-\alpha \left( \hat{z}\times \mathbf{k}\right)\cdot \sigma +\beta D\left( k_{x}^{2}-k_{y}^{2}\right) \mathbf{n}%
\cdot \mathbf{\sigma } ,
\end{eqnarray}
where $\mathbf{k}=(k_x,k_y,0)$ and $\mathbf{\sigma }=(\sigma_x,\sigma_y,\sigma_z)$ is the vector of Pauli matrix in spin space. In the above equation, the first term represents the kinetic energy, the second signifies the Rashba SOC (where $\alpha$ quantifies its strength), and the last one corresponds to the altermagnetic term---characterized by the combination of four-fold rotation and time-reversal symmetry ($C_4\mathcal{T}$). In the altermagnetic term, $\beta<1$ must be satisfied to ensure an elliptical Fermi surface (rather than hyperbolic), and the N\'{e}el vector is defined as $\mathbf{n}=(n_x,\;n_y,\;n_z)=( \sin \theta _n\cos \varphi _n,\sin \theta _n\sin\varphi _n,\cos \theta _n)$, where $\theta _n$ and $\varphi _n$ denote the polar and azimuthal angles, respectively. Here, the consideration of a N\'{e}el vector $\mathbf{n}$ with an arbitrary orientation is motivated by two key factors. First, experiments have demonstrated that the orientation of the N\'{e}el vector can be electrically probed and controlled \cite{Mn5Si3_1,neelprobe}, making its deliberate variation both relevant and experimentally accessible. Second, according to Ref. \cite{altermagnet3}, in both predicted and realized 2D $d$-wave altermagnets, the N\'{e}el vector orientation can be either perpendicular to the 2D plane (out-of-plane) or lie within the plane (in-plane). Specifically, the N\'{e}el vector is out-of-plane in materials such as ${\rm FeX}$ (${\rm X = S, Se}$) \cite{altermagnet6}, the ${\rm Ca(FeN)_2}$ monolayer \cite{out1}, ${\rm Fe_2WTe_4}$ \cite{out2}, ${\rm V_2Te_2O}$ \cite{out4}, ${\rm Cr_2Te_2O}$ monolayer \cite{out5}, and ${\rm Fe_2Se_2O}$ monolayer \cite{out6}. In contrast, the N\'{e}el vector is in-plane in materials such as ${\rm SrRuO_3}$ thin films \cite{opt2}, the ${\rm CrO}$ monolayer \cite{in1}, and ${\rm RuF_4}$ \cite{in3}. Therefore, to maintain generality and encompass the full range of possible configurations---whether intrinsic or electrically tuned---we consider $\mathbf{n}$ with an arbitrary orientation. On the other hand, the Rashba term tends to favor in-plane spin orientations. This is because the substrate or gate voltage breaks the inversion symmetry of the altermagnet, causing moving electrons to experience an effective transverse magnetic field. This transverse field can only induce Rashba SOC with in-plane spins, as described in Refs. \cite{alterRKKY3,subs}. By diagonalizing the Hamiltonian of Eq. (\ref{eq1}), the energy dispersion can be solved as
\begin{eqnarray}\label{eq2}
\epsilon_{\pm}=Dk^{2} \pm \epsilon_{0},
\end{eqnarray}
with
\begin{eqnarray}\label{eq3}
\epsilon_{0}=\sqrt{ \beta ^{2}D^{2}\left( k_{x}^{2}-k_{y}^{2}\right) ^{2}+\alpha
^{2}k^{2}-2\alpha \beta D\left( k_{x}^{2}-k_{y}^{2}\right) \left(\mathbf{k}\times\mathbf{n}\right)_z}.
\end{eqnarray}
\par
In the absence of SOC, the electron spin is a conserved quantity since the altermagnetic term contributes only nonrelativistic effects. In this case, the Fermi surfaces are standardized ellipses, and the spins on different ellipses can be easily classified. To demonstrate this point, the Fermi surfaces and the spin expectation $s_{i}(\epsilon_\pm)=\langle \Psi_{\pm}|\hbar\sigma _{i}/2|\Psi_{\pm}\rangle$ ($i=x,y,z$ and $\Psi_{\pm}$ are the eigenstates of Hamiltonian $H$) are plotted in Fig. 1(a). Apparently, only two solid colors---red and blue---are present here. They correspond respectively to spin-up ($s_{z}=+1$) and spin-down ($s_{z}=-1$) states along the direction of the N\'{e}el vector, spatially segregated on distinct elliptical Fermi surfaces. This describes a novel type of spin splitting, which distinguishes it from the ferromagnetic case and is a direct manifestation of altermagnetism. It is worth noting that altermagnetism persists throughout momentum space, except along the lines $k_x=\pm k_y$ where its vanishing enables the existence of nodal lines.
\par
For the introduction of weak SOC, we assume that the N\'{e}el vector orientation remains unmodified, an assumption consistent with Refs. \cite{alterRKKY2,alterRKKY3,subs}, since the SOC strength selected in our work is of the same order of magnitude as that in Ref. \cite{alterRKKY3}. When SOC is present, two significant effects occur: (1) The shape of the band undergoes substantial modification. Most notably, the nodal lines at $k_x=\pm k_y$ are fully gapped, resulting in non-elliptical Fermi surfaces, as shown in Fig. 1(b); (2) The distribution of the spin expectation is altered. For comparison, we plot only the dominant spin expectation (i.e., N\'{e}el-vector-aligned spin expectation) in Figs. 1(b) and 1(c). Comparing Fig. 1(b) with Fig. 1(a), one can find that the colors representing spin expectation on the Fermi surface are no longer pure red or blue. Instead, the hues become diluted. This indicates that the original altermagnetism-induced spin expectation is suppressed, and partially replaced by its SOC-determined counterpart. Notably, these effects are direction-dependent and sensitive to N\'{e}el-vector orientation. Specifically, for the case of $\mathbf{n}=(0,0,1)$, the strongest suppression of $s_z$ occurs near $k_x=\pm k_y$, where $s_z$ nearly vanishes [the white colour in Fig. 1(b)]. For the N\'{e}el vector with an in-plane orientation, the dominant spin expectation is changed to $s_x$. In this case, not only are Fermi surfaces further distorted (breaking $C_4\mathcal{T}$ symmetry), but the location of maximal spin-expectation suppression also shifts [Fig. 1(c)].
\par
Overall, compared to the system with only altermagnetism, the band shape and the spin expectation distribution are significantly changed as SOC is present. Given that the RKKY interaction is highly sensitive to band structure and electron spin, this raises the questions: how the RKKY interaction behaves in the system where altermagnetism and SOC coexist, and whether the N\'{e}el vector orientation can be reflected by the RKKY interaction.
\par
To construct a model for the RKKY interaction, two magnetic impurities are assumed to be placed on the 2D altermagnet, where one impurity is located at $\mathbf{r}_1$ and the other is at $\mathbf{r}_2$. As impurities are surrounded by itinerant electrons of the host material, each of them would interact with electrons through a contact interaction described by $H_{i}=J\mathbf{S}_{i}\cdot \mathbf{\sigma}\delta(\mathbf{r}-\mathbf{r}_i)$, with $\mathbf{S}_{i}$ ($i=1$, $2$) denoting the spin of the impurity. Due to multiple scattering of itinerant electrons between the two impurities, an effective indirect interaction (i.e., RKKY interaction) between impurities is mediated. By retaining terms up to second order in $J$ and following the standard perturbation theory \cite{pertub1,pertub2,pertub3,pertub4}, the RKKY interaction can be expressed as
\begin{eqnarray}\label{eq4}
H_{R}=-\frac{J^{2}}{\pi }{\rm Im}\int_{-\infty }^{u_F}d\omega {\rm Tr}%
\left[ \left( S_{1}\cdot \sigma \right) G\left( \mathbf{R},\omega \right)
\left( S_{2}\cdot \sigma \right) G\left( -\mathbf{R},\omega \right) \right],
\end{eqnarray}
where zero temperature is assumed and $G\left( \mathbf{R},\omega \right)$ ($\mathbf{R}=\mathbf{r}_1-\mathbf{r}_2$) is the retarded Green's function.
\par
Before evaluating the RKKY interaction, the retarded Green's function in the real space has to be obtained, which can be calculated as
\begin{equation}\label{eq5}
G\left( \omega,\pm \mathbf{R} \right)=\frac{1}{\left(2\pi\right)^2}\int d\mathbf{k}\frac{1}{\omega+i0^+-H}e^{\pm i\mathbf{k}\mathbf{R}}.
\end{equation}
Since no approximation is made for the SOC strength $\alpha$, numerical calculation is essential to compute the above Green's function. Despite this, by inserting $H$ of Eq. (\ref{eq1}) into the above equation, one can still simplify $G\left( \omega,\pm \mathbf{R} \right)$ to the following form
\begin{eqnarray}\label{eq6}
G\left( \pm \mathbf{R},\omega \right) =g_{0}^{\pm }\sigma _{0}+g_{n}^{\pm }%
\mathbf{n}\cdot \mathbf{\sigma +}g_{x}^{\pm }\sigma _{x}+g_{y}^{\pm }\sigma
_{y},
\end{eqnarray}
with
\begin{eqnarray}\label{eq7}
\begin{split}
g_{0}^{\pm }&=\frac{1}{\left( 2\pi \right) ^{2}}\int d\mathbf{k}\frac{%
\left( \omega -Dk^{2}\right) }{\left( \omega -Dk^{2}\right) ^{2}-\epsilon
^{2}}e^{\pm i\mathbf{kR}}, \\
g_{n}^{\pm }&=\frac{1}{\left( 2\pi \right) ^{2}}\int d\mathbf{k}%
\frac{\beta D\left( k_{x}^{2}-k_{y}^{2}\right) }{\left( \omega -Dk^{2}\right)
^{2}-\epsilon ^{2}}e^{\pm i\mathbf{kR}},\\
g_{x}^{\pm }&=\frac{1 }{\left( 2\pi \right) ^{2}}\int d\mathbf{k}%
\frac{\alpha k_{y}}{\left( \omega -Dk^{2}\right) ^{2}-\epsilon ^{2}}e^{\pm i\mathbf{%
kR}}, \\
g_{y}^{\pm }&=\frac{1 }{\left( 2\pi \right) ^{2}}\int d\mathbf{k}%
\frac{-\alpha k_{x}}{\left( \omega -Dk^{2}\right) ^{2}-\epsilon ^{2}}e^{\pm i\mathbf{%
kR}},
\end{split}
\end{eqnarray}
where $g_{j}^{\pm }$ ($j=0$, $n$, $x$, $y$) is shorthand for $g_{j}\left(\pm \mathbf{R}\right)$.
\par
By plugging Eq. (\ref{eq6}) into Eq. (\ref{eq4}) and taking the trace over the spin degrees of freedom, the RKKY interaction $H_{R}$ can be further simplified as
\begin{eqnarray}\label{eq8}
H_{R}=\sum\limits_{i=x,y,z}\left[ J_{ii}S_{1}^{i}S_{2}^{i}+J_{DM}^{i}%
\left( \mathbf{S}_{1}\times \mathbf{S}_{2}\right) _{i}+J_{f}^{i}\left(
S_{1}^{j}S_{2}^{k}+S_{1}^{k}S_{2}^{j}\right) \right].
\end{eqnarray}
where ($j$, $k$) form an even permutation for the Levi-Civita symbol with a fixed index $i$. In the above equation, the first term in square brackets describes the collinear components of the RKKY interaction, while the last two terms denote the noncollinear components, representing the DM terms and frustrated terms respectively. The magnitudes of the collinear components are given by
\begin{eqnarray}\label{eq9}
\begin{split}
J_{xx}&=J_{H}-\frac{2J^{2}}{\pi }{\rm Im}\int\nolimits_{-\infty
}^{u_F}\left(g_{x}^{+}g_{x}^{-}-g_{y}^{+}g_{y}^{-}+2n_{x}f_{x}+2n_{x}^{2}g_{n}^{+}g_{n}^{-}\right) d\omega,\\
J_{yy}&=J_{H}-\frac{2J^{2}}{\pi }{\rm Im}\int\nolimits_{-\infty
}^{u_F}\left(g_{y}^{+}g_{y}^{-}-g_{x}^{+}g_{x}^{-}+2n_{y}f_{y}+2n_{y}^{2}g_{n}^{+}g_{n}^{-}\right) d\omega, \\
J_{zz}&=J_{H}-\frac{2J^2}{\pi }{\rm Im}\int\nolimits_{-\infty
}^{u_F}\left(2n_{z}^{2}g_{n}^{+}g_{n}^{-}-g_{y}^{+}g_{y}^{-}-g_{x}^{+}g_{x}^{-}\right)
d\omega, \\
\end{split}
\end{eqnarray}
with
\begin{eqnarray} \label{eq10}
\begin{split}
 J_{H}&=-\frac{2J^{2}}{\pi }{\rm Im}\int\nolimits_{-%
\infty }^{u_F}\left(g_{0}^{+}g_{0}^{-}-g_{n}^{+}g_{n}^{-}-n_{x}f_{x}-n_{y}f_{y}\right) d\omega .
\end{split}
\end{eqnarray}
where $f_{i}=g_{n}^{+}g_{i}^{-}+g_{n}^{-}g_{i}^{+}$. For the noncollinear components, their magnitudes are given by
\begin{eqnarray}\label{eq11}
\begin{split}
J_{DM}^{x}&=-\frac{2J^{2}}{\pi }{\rm Im}\int\nolimits_{-\infty }^{u_F}i\left[
n_{x}\left( g_{0}^{+}g_{n}^{-}-g_{0}^{-}g_{n}^{+}\right)
+g_{0}^{+}g_{x}^{-}-g_{0}^{-}g_{x}^{+}\right] d\omega , \\
J_{DM}^{y}&=-\frac{2J^{2}}{\pi }{\rm Im}\int\nolimits_{-\infty }^{u_F}i\left[
n_{y}\left( g_{0}^{+}g_{n}^{-}-g_{0}^{-}g_{n}^{+}\right)
+g_{0}^{+}g_{y}^{-}-g_{0}^{-}g_{y}^{+}\right] d\omega , \\
J_{DM}^{z}&=-\frac{2J^{2}}{\pi }{\rm Im}\int\nolimits_{-\infty
}^{u_F}in_{z}\left( g_{0}^{+}g_{n}^{-}-g_{0}^{-}g_{n}^{+}\right) d\omega, \\
J_{f}^{x}&=-\frac{2J^{2}}{\pi }{\rm Im}\int\nolimits_{-\infty
}^{u_F}n_{z}\left(g_{n}^{+}g_{y}^{-}+g_{n}^{-}g_{y}^{+}+2n_{y}g_{n}^{+}g_{n}^{-}\right)
d\omega, \\
J_{f}^{y}&=-\frac{2J^{2}}{\pi }{\rm Im}\int\nolimits_{-\infty
}^{u_F}n_{z}\left(g_{n}^{+}g_{x}^{-}+g_{n}^{-}g_{x}^{+}+2n_{x}g_{n}^{+}g_{n}^{-}\right)
d\omega , \\
J_{f}^{z}&=-\frac{2J^{2}}{\pi }{\rm Im}\int\nolimits_{-\infty }^{u_F}\left(
n_{x}f_{y}+n_{y}f_{x}+2n_{x}n_{y}g_{n}^{+}g_{n}^{-}+f_{xy}\right) d\omega,
\end{split}
\end{eqnarray}
where $f_{xy}=g_{x}^{+}g_{y}^{-}+g_{x}^{-}g_{y}^{+}$.

\section{Results and Discussion}
\begin{table}[th]
\caption{A breakdown of the nonzero RKKY components of 2D electron gas systems. Finite Fermi energy is considered. $\theta_R=\arctan(R_y/R_x)$, denoting the azimuthal angle of the inter-impurity vector.}
\centering
\label{table1}
\begin{spacing}{2}
\begin{tblr}{
  hlines, vlines,
  colspec = {Q[c, wd=0.4cm] Q[c, wd=2.08cm] Q[c, wd=2.45cm] Q[c, wd=2.25cm]},
  row{1} = {2-4}{c},
  cell{1}{1} = {r=2}{c},
  cell{1}{2} = {r=2}{c},
  cell{1}{3} = {c=2}{c},
  rows = {ht=0.5cm, abovesep=1pt, belowsep=1pt},
  row{1} = {ht=0.1cm, abovesep=0pt, belowsep=0pt},
  row{2} = {ht=0.1cm, abovesep=0pt, belowsep=0pt},
}
$\theta_R$ & {collinear\\[-10pt]components} & \SetCell[c=2]{c} {noncollinear\\[-10pt]components} &  \\
          &                            & DM terms & frustrated terms \\
$0$       & {$J_{xx}=J_{zz}\neq J_{yy}$ \\ [-10pt](XYX)} & $J^{y}_{DM}$ & -- \\
$\pi/2$   & {$J_{xx}\neq J_{yy}=J_{zz}$ \\ [-10pt](XYY)} & $J^{x}_{DM}$ & -- \\
$\pi/4$   & {$J_{xx}=J_{yy}\neq J_{zz}$ \\[-10pt] (XXZ)} & $J^{x}_{DM}=-J^{y}_{DM}\neq0$ & $J^{z}_{f}$ \\
\end{tblr}
\end{spacing}
\end{table}

In this section, the discussion is structured in two parts: (1) The effect of the N\'{e}el vector on the RKKY interaction; (2) The effect of altermagnetism-SOC coexistence on the RKKY interaction. For later comparative analysis, the RKKY behavior mediated by 2D electron gas \cite{electrongas}---with Hamiltonian $H$ given in Eq. (1) for $\beta=0$---is compiled in Table \ref{table1}. Additionally, we confine the N\'{e}el vector to the $zx$-plane, as this configuration provides an instructive example to reveal the connection between the RKKY interaction and the N\'{e}el vector.

\subsection{Effect of the N\'{e}el vector on RKKY}
\par
\subsubsection{In the absence of SOC}
Firstly, we consider the RKKY interaction in the absence of SOC (i.e., $\alpha=0$). In this scenario, the terms $g_{x}^\pm$ and $g_{y}^\pm$ vanish, and $g_{\eta}^+=g_{\eta}^-$ ($\eta=0,n$) is guaranteed by the preserved inversion symmetry. As a result, the RKKY interaction of Eq. (\ref{eq8}) can be further simplified as
\begin{align}
H_{R}&=J_H \mathbf{S}_{1}\cdot \mathbf{S}_{2}+ J_I\sum\limits_{i=x,y,z} n_i^2S_{1}^{i}S_{2}^{i}+ \sum\limits_{i=x,y,z} J_{f}^i\left(
S_{1}^{j}S_{2}^{k}+S_{1}^{k}S_{2}^{j}\right), \label{eq12} \\
&=J_H \mathbf{S}_{1}\cdot \mathbf{S}_{2}+J_I\left(\mathbf{n}\cdot \mathbf{S}_1\right)\left(\mathbf{n}\cdot \mathbf{S}_2\right) , \label{eq13}
\end{align}
with $J_f^i=n_jn_kJ_{I}$ and
\begin{eqnarray}\label{eq14}
\begin{split}
J_{H}&=-\frac{J^{2}}{2\pi ^{2}\left( 1-\beta ^{2}\right) D}\frac{\sin \left[
\sqrt{\frac{u_{F}}{D}}\left( R_{+}+R_{-}\right) \right] }{\sqrt{R_{+}R_{-}}%
\left( R_{+}+R_{-}\right) }, \\
J_{I}&=-\frac{J^{2}}{8\pi ^{2}\left( 1-\beta ^{2}\right) D}%
\sum\limits_{s=\pm }\frac{\sin \left( 2R_{s}\sqrt{\frac{u_{F}}{D}}\right) }{%
R_{s}^{2}}-J_{H}, \\
\end{split}
\end{eqnarray}
where $R_{s}=\sqrt{R_{x}^{2}/\left( 1+s\beta \right) +R_{y}^{2}/\left( 1-s\beta
\right) }$ and ($j$, $k$) constituting an even permutation of the indices for the Levi-Civita symbol when $i$ is fixed. The expressions in Eq. (\ref{eq14}) give the asymptotic results, obtained by using the approximation of $R_s\sqrt{u_F/D}\gg 1$.
\par
The final term in Eq. (\ref{eq12}) corresponds to the frustrated term. This term combines with the second term to form the Ising interaction $J_I$ in Eq. (\ref{eq13}), whose direction is entirely governed by the N\'{e}el vector $\mathbf{n}$. For example, when $\mathbf{n}=(0,0,1)$ (i.e., $\theta_n=0$), only the collinear components---the Heisenberg term $J_H\mathbf{S}_{1}\cdot \mathbf{S}_{2}$ and the Ising term $J_IS_{1}^{z}S_{2}^{z}$---survive. The resulting RKKY interaction exhibits anisotropy due to the Ising term, which corresponds to a $z$-axis collinear component and naturally yields an XXZ spin model. This agrees with the results in Refs. \cite{alterRKKY1,alterRKKY2,alterRKKY3}, where the N\'{e}el vector is fixed along the $z$-axis. Owing to the directional consistency [Eq. (\ref{eq13})] between the Ising term and the N\'{e}el vector, the N\'{e}el vector orientation can be directly ascertained by detecting the Ising term. Note that this consistency holds if and only if $\theta_R\neq (2m+1)\pi/4$ ($m\in\mathbb{Z}$), where $\theta_R=\arctan(R_y/R_x)$ is the azimuthal angle of the inter-impurity vector. For impurities aligned along $\theta_R=(2m+1)\pi/4$, the RKKY spin model reduces to the XXX model for any $\theta_n$, as evident from Eq. (\ref{eq14}), where $R_x^2=R_y^2$ forces $J_I=0$. Equivalently, $J_I=0$ can also be attributed to the complete annihilation of altermagnetism, i.e., the vanishing spin expectation in the direction of $\theta=(2m+1)\pi/4$ ($k_x=\pm k_y$) [Fig. 1(b)]. Here, only the kinetic term $Dk^2\sigma_0$ remains, naturally yielding an isotropic XXX spin model ($J_I=0$). Thus, to probe the N\'{e}el vector orientation via the Ising term in altermagnets with absent or negligible SOC, impurity configurations with $\theta_R=(2m+1)\pi/4$ should be avoided.
\par

\subsubsection{In the presence of SOC}
\begin{table}[th]
\caption{A breakdown of the nonzero RKKY components of 2D $d$-wave altermagnets in the presence of Rashba SOC. Finite Fermi energy is considered. With $\varphi_n = 0$ fixed, $\theta_n$ is the polar angle of the N\'{e}el vector, and $\theta_R$ is the azimuthal angle of the inter-impurity vector.}
\centering
\label{table2}
\begin{spacing}{2}
\begin{tblr}{
  hlines, vlines,
  colspec = {Q[c, wd=0.4cm] Q[c, wd=0.4cm] Q[c, wd=1.6cm] Q[c, wd=2.15cm] Q[c, wd=2.25cm]},
  row{1} = {2-5}{c},
  cell{1}{1} = {r=2}{c},
  cell{1}{2} = {r=2}{c},
  cell{1}{3} = {r=2}{c},
  cell{1}{4} = {c=2}{c},
}
$\theta_R$ & $\theta_n$ & {collinear \\ [-10pt] components} & \SetCell[c=2]{c} {noncollinear \\ [-10pt] components} &  \\
          &            &                            & DM terms & frustrated terms \\
\SetCell[r=3]{c} $0$     & $0$        & \SetCell[r=3]{c} $J_{xx}$, $J_{yy}$, $J_{zz}$ & $J^{y}_{DM}$ & {-} \\
                         & $\pi/2$    &                                             & $J^{y}_{DM}$ & {-} \\
                         & $\pi/3$    &                                             & $J^{y}_{DM}$ & $J^{y}_{f}$ \\
\SetCell[r=3]{c} $\pi/2$ & $0$        & \SetCell[r=3]{c} $J_{xx}$, $J_{yy}$, $J_{zz}$ & $J^{x}_{DM}$ & {-} \\
                         & $\pi/2$    &                                             & $J^{x}_{DM}$ & {-} \\
                         & $\pi/3$    &                                             & {$J^{x}_{DM}$, $J^{z}_{DM}$} & $J^{y}_{f}$ \\
\SetCell[r=3]{c} $\pi/4$ & $0$        & \SetCell[r=3]{c} $J_{xx}$, $J_{yy}$, $J_{zz}$ & {$J^{x}_{DM}$, $J^{y}_{DM}$ \\ [-10pt]($J^{y}_{DM}=-J^{x}_{DM})$} & $J^{z}_{f}$ \\
                         & $\pi/2$    &                                             & {$J^{x}_{DM}$, $J^{y}_{DM}$ \\ [-10pt]($J^{y}_{DM}\neq -J^{x}_{DM})$} & $J^{z}_f$ \\
                         & $\pi/3$    &                                             & {$J^{x}_{DM}$, $J^{y}_{DM}$, $J^{z}_{DM}$ \\ [-10pt]($J^{y}_{DM}\neq -J^{x}_{DM})$} & {$J^{x}_{f}$, $J^{y}_{f}$, $J^{z}_f$} \\
\end{tblr}
\end{spacing}
\end{table}
Once SOC is present, the RKKY interaction exhibits increased complexity. Specifically, SOC disrupts the original RKKY spin model, making it strongly dependent on the SOC strength $\alpha$ (analyzed later in Subsection B-2). This implies that the N\'{e}el vector's effect on the RKKY spin model varies with $\alpha$, resulting in significantly greater difficulty in ascertaining the N\'{e}el vector orientation from the Ising term.
\par
To find another way to identify the general effect of the N\'{e}el vector on the RKKY interaction---ensuring that this effect is independent of the specific value of $\alpha$ ($\alpha \neq 0$)---we list the nonzero RKKY components of altermagnets for different impurity configurations in Table \ref{table2}. The survival of these components remains unchanged for any nonzero $\alpha$, as numerically verified via Eqs. \eqref{eq7}, \eqref{eq9}--\eqref{eq11}. According to Table \ref{table2}, one can find that at $\theta_R=\pi/4$---identified as the optimal impurity configuration---the DM terms most effectively reveal the N\'{e}el vector's influence. In particular, for $\mathbf{n}=(0,0,1)$ (i.e., $\theta_n=0$), only the in-plane components of the DM interaction survive. These components exhibit antisymmetry: $J_{DM}^{x}=-J_{DM}^{y}$, protected by the $C_4\mathcal{T}$ symmetry of the band structure [Fig. 1(b)]. Conversely, for in-plane orientations of $\mathbf{n}$ (e.g., $\theta_n=\pi/2$), this antisymmetry is broken ($J_{DM}^{x}\neq-J_{DM}^{y}$) due to the loss of $C_4\mathcal{T}$ symmetry [Fig. 1(c)]. At intermediate orientations of $\mathbf{n}$ (e.g., $\theta_n=\pi/3$), along with persistent antisymmetry breaking, a new DM component $J_{DM}^z$ emerges. Overall, DM terms' high sensitivity to the N\'{e}el vector orientation still enables qualitative estimation of its orientation through DM-term measurements.

\subsection{Effects of altermagnetism-SOC coexistence on RKKY}

\subsubsection{\textbf{Synergistic effects}: generating new DM component $J_{DM}^z$ }
Table \ref{table2} reveals an unusual phenomenon: the emergence of the out-of-plane component $J_{DM}^z$ for the DM interaction. Crucially, $J_{DM}^z$ vanishes in systems with only altermagnetism or only SOC (Table \ref{table1}), indicating that its nonzero value requires the coexistence of both. In other words, altermagnetism and SOC synergistically generate a new DM component $J_{DM}^z$. Table \ref{table2} further shows that this synergy functions only for specific N\'{e}el vector orientations $\mathbf{n}$ with both out-of-plane ($n_z$) and in-plane ($n_x$ or $n_y$) components. This situation contrasts significantly with previous studies \cite{alterRKKY1,alterRKKY2,alterRKKY3}, where $J_{DM}^z$ is absent due to first-order SOC expansion and purely axial N\'{e}el vector orientation with $\mathbf{n}=(0,0,1)$. To understand the related physics, one can review Eqs. (\ref{eq3}) and (\ref{eq11}). It is found that the nonzero $J_{DM}^z$ is determined by two factors: (i) The in-plane component ($n_x$ or $n_y$) of the N\'{e}el vector enters the band energy and combines with SOC to break the $C_4\mathcal{T}$ symmetry of the band [Fig. 1(c)]. Consequently, the inversion symmetry of the relevant Green's functions is broken, i.e., $g_{0,n}^+\neq g_{0,n}^-$, leading to $g_0^+g_n^-\neq g_0^-g_n^+$; (ii) The out-of-plane component $n_z$ of $\mathbf{n}$ couples to the altermagnetism-induced component $g^{\pm}_n$ of the Green's function. The combination of these two factors yields a nonzero integrand $n_z(g_0^+g_n^--g_0^-g_n^+)$ in $J_{DM}^z$ [Eq. (\ref{eq11})], leading to $J_{DM}^z\neq0$.
\par
\subsubsection{\textbf{Multifaceted effects}:  dictating RKKY spin models}
\begin{figure*}[!htb]
\centering \includegraphics[width=0.7\textwidth]{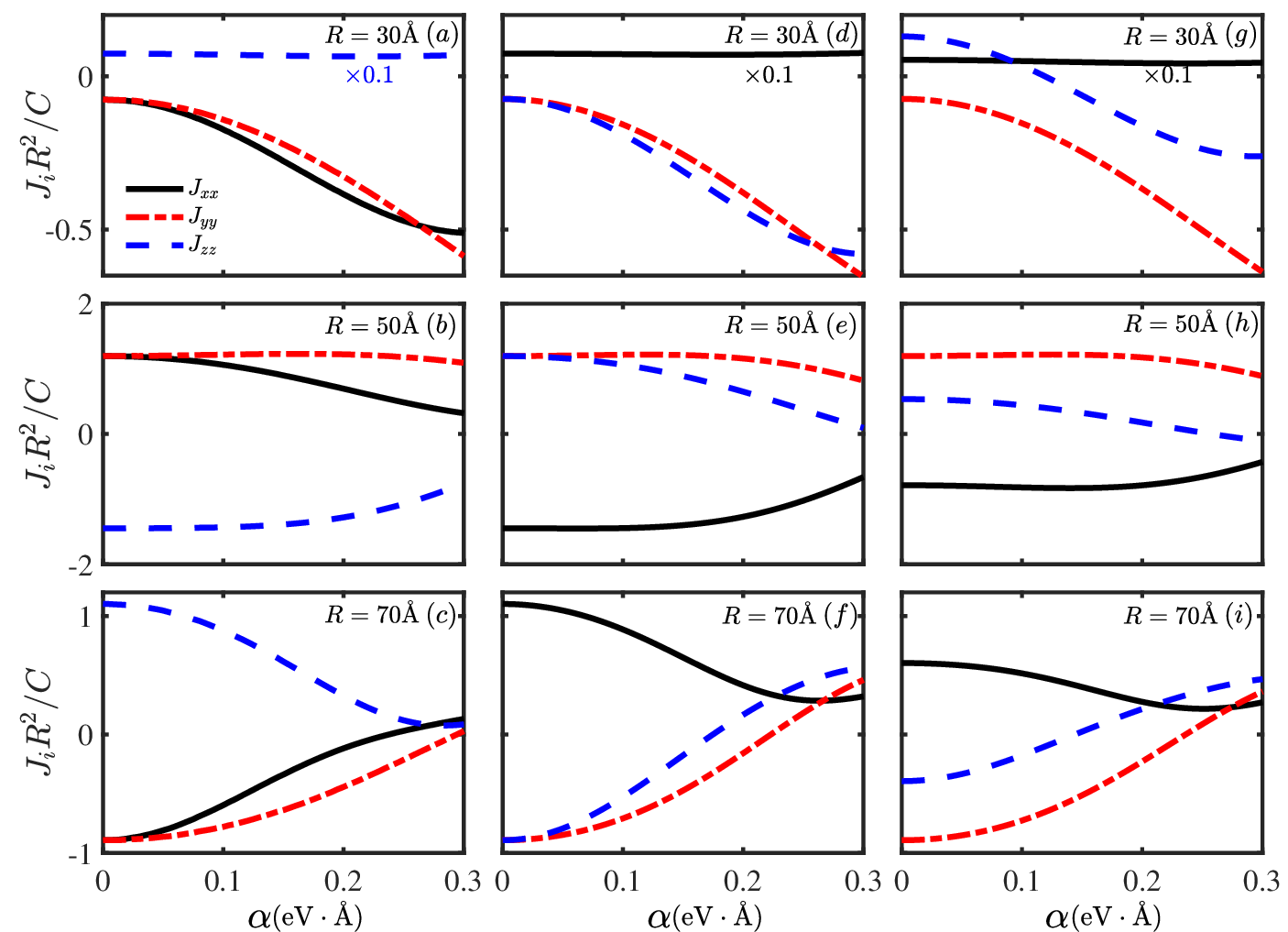}
\caption{(Color online) The RKKY components $J_{ii}$ ($i=x,y,z$) versus the SOC strength $\alpha$ for different N\'{e}el vector orientations (fixed $\varphi_n=0$): (a-c) $\theta_n=0$; (d-f) $\theta_n=\pi/2$; (g-i) $\theta_n=\pi/3$. Impurities are aligned along $\theta_R=0$ with varying distances (see annotated values in the above figure). Other parameters are identical to Fig. 1(b).}
\end{figure*}

\begin{figure*}[!htb]
\centering \includegraphics[width=0.7\textwidth]{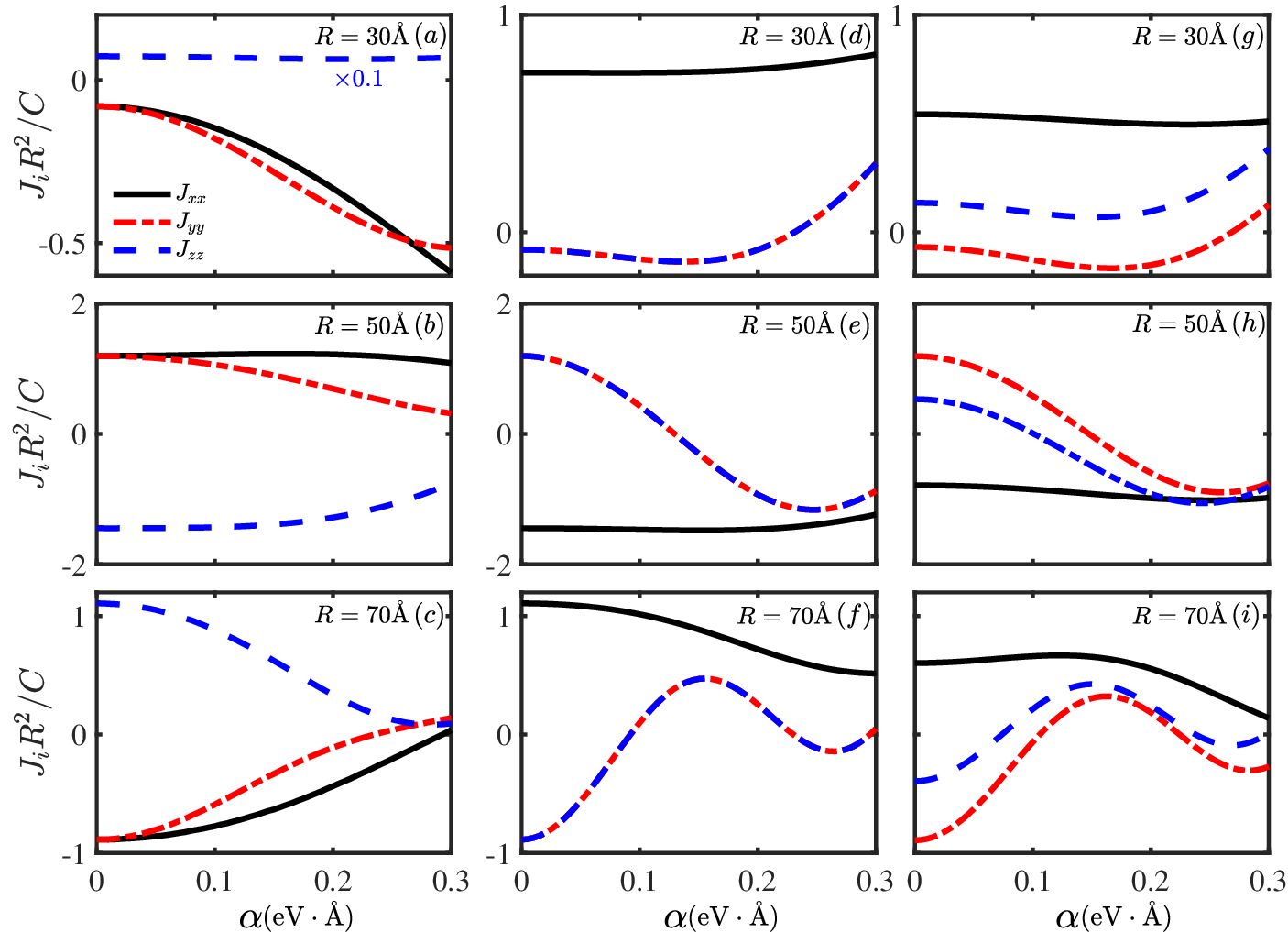}
\caption{(Color online) The RKKY components $J_{ii}$ ($i=x,y,z$) versus the SOC strength $\alpha$ for different N\'{e}el vector orientations (fixed $\varphi_n=0$): (a-c) $\theta_n=0$; (d-f) $\theta_n=\pi/2$; (g-i) $\theta_n=\pi/3$. Impurities are aligned along $\theta_R=\pi/2$ with varying distances (see annotated values in the above figure). Other parameters are identical to Fig. 2.}
\end{figure*}

\begin{figure*}[!htb]
\centering \includegraphics[width=0.7\textwidth]{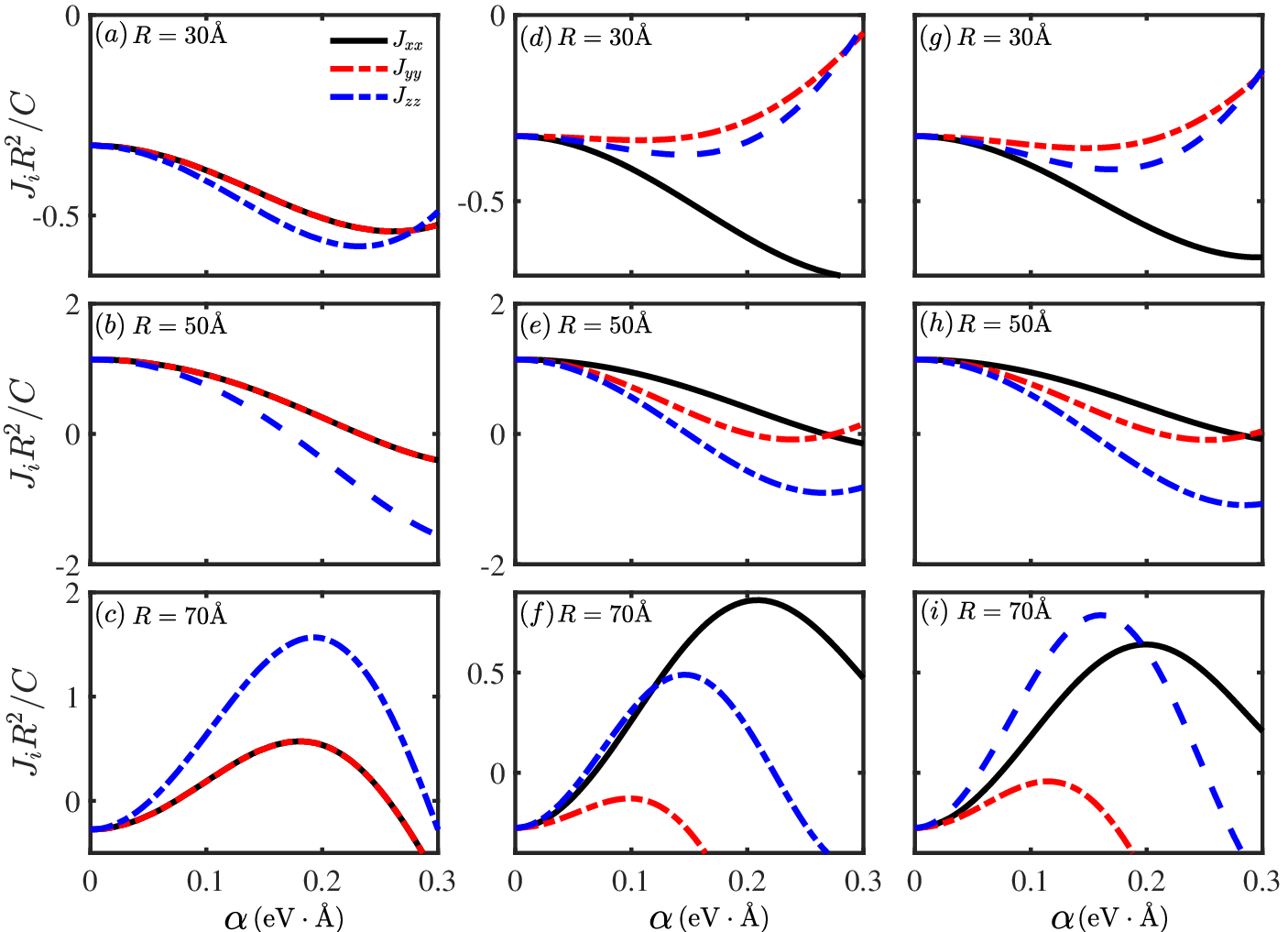}
\caption{(Color online) The RKKY components $J_{ii}$ ($i=x,y,z$) versus the SOC strength $\alpha$ for different N\'{e}el vector orientations (fixed $\varphi_n=0$): (a-c) $\theta_n=0$; (d-f) $\theta_n=\pi/2$; (g-i) $\theta_n=\pi/3$. Impurities are aligned along $\theta_R=\pi/4$ with varying distances (see annotated values in the above figure). Other parameters are identical to Fig. 2.}
\end{figure*}

Different from the single effect on the DM term $J^z_{DM}$, altermagnetism-SOC coexistence induces multifaceted effects on the RKKY spin models as the SOC strength $\alpha$ varies. To demonstrate this, we plot $\alpha$-dependent RKKY components $J_{ii}$, as shown in Figs. 2-4. Given the direction dependence and N\'{e}el vector sensitivity of SOC-induced effects on the band (as discussed in page 3), we study the RKKY spin model across diverse impurity configurations and N\'{e}el vector orientations. Based on these plots, the evolution of the RKKY spin model is summarized in Table \ref{table3}, where we identify five distinct mechanisms governing the spin model, as follows:
\begin{itemize}
 \item[\safebigding{172}]  Altermagnetism dominates the spin model;
  \item[\safebigding{173}] SOC dominates the spin model;
 \item[ \safebigding{174}] The kinetic term dominates the spin model;
  \item[\safebigding{175}] The spin model results from competition between altermagnetism and SOC;
  \item[\safebigding{176}] Altermagnetism and SOC coincidentally produce the same spin model.
\end{itemize}
 The former three mechanisms describe the effect arising from a single term in the Hamiltonian of Eq. (\ref{eq1}), whereas the latter two correspond to the combined effect of altermagnetism and SOC.
\begin{table}[th]
\caption{Evolution of spin models with increasing $\alpha$ (from $\alpha=0$). Finite Fermi energy $u_F$ is considered. \safebigding[1.2]{172}$\sim$\safebigding[1.2]{176} are the five fundamental mechanisms listed in Sec. III-B-2.}
\centering
\label{table3}
\begin{spacing}{2}
\begin{tblr}{
  hlines, vlines,
  colspec = {Q[c, wd=1cm] Q[c, wd=1cm] Q[c, wd=5cm]},
}
$\theta_R$ & $\theta_n$ & spin models $\left(\alpha\!: 0 \rightarrow \text{large}\right)$ \\
\SetCell[r=3]{c} $0$ & $0$ & ${\color{mymagenta}\uuline{\textcolor{myblack}{{\rm XXZ}\; \safebigding[1.2]{172}\rightarrow {\rm XYZ}\;  \safebigding[1.2]{175}}}}$ \\
                    & $\pi/2$ & XYY $\safebigding[1.2]{172}$ $\rightarrow$XYZ $ \safebigding[1.2]{175}$ \\
                    & $\pi/3$ & ${\color{myblue}\uuline{\textcolor{myblack}{{\rm XYZ}\;\safebigding[1.2]{172} }}}$ \\
\SetCell[r=3]{c} $\pi/2$ & $0$ & XXZ $\safebigding[1.2]{172}$$\rightarrow$XYZ $ \safebigding[1.2]{175}$ \\
                         & $\pi/2$ & ${\color{myred}\uuline{\textcolor{myblack}{{\rm XYY}\;\safebigding[1.2]{176}}}}$ \\
                         & $\pi/3$ & XYZ $\safebigding[1.2]{172}$ \\
\SetCell[r=3]{c} $\pi/4$ & $0$ & XXX $\safebigding[1.2]{174}$$\rightarrow$XXZ $ \safebigding[1.2]{173}$ \\
                         & $\pi/2$ & \SetCell[r=2]{c} XXX $\safebigding[1.2]{174}$$\rightarrow$ XYZ  $\safebigding[1.2]{175}$ \\
                         & $\pi/3$ & \\
\end{tblr}
\end{spacing}
\end{table}
\par
To facilitate the explanation of the spin model evolution, impurity configurations are divided into two categories: impurities aligned along the axial (e.g., $\theta_R=0$ and $\pi/2$) or non-axial (e.g., $\theta_R=\pi/4$) directions. For the former configuration, Table \ref{table3} shows three representative evolutions:
\par
(1) XXZ $\rightarrow$ XYZ (magenta double underline in Table \ref{table3}). For $\theta_n=0$ (i.e., the N\'{e}el vector points along the $z$-axis), the system still preserves $C_4\mathcal{T}$ symmetry, which leads to
\begin{eqnarray}\label{eq15}
\begin{split}
g_\eta^+&=g_\eta^-, \\
g_i^+&=-g_i^-,
\end{split}
\end{eqnarray}
where $\eta=0$, $n$ and $i=x$, $y$, $z$. For impurities aligned at $\theta_R=0$ (i.e., along the $x$-axis), we further obtain $g^\pm_x=0$ since the integrand in Eq. (\ref{eq7}) is an odd function of $k_y$. Consequently, the collinear components $J_{ii}$ in Eq. (\ref{eq9}) reduce to
\begin{eqnarray}\label{eq16}
\begin{split}
J_{xx}&=\frac{2J^{2}}{\pi }{\rm Im}\int\nolimits_{-\infty }^{u_F}\left(
-g_{0}^{+2}+g_{n}^{+2}-g_{y}^{+2}\right) d\omega , \\
J_{yy}&=\frac{2J^{2}}{\pi }{\rm Im}\int\nolimits_{-\infty }^{u_F}\left(
-g_{0}^{+2}+g_{n}^{+2}+g_{y}^{+2}\right) d\omega, \\
J_{zz}& =\frac{2J^{2}}{\pi }{\rm Im}\int\nolimits_{-\infty }^{u_F} \left(
-g_{0}^{+2}-g_{n}^{+2}-g_{y}^{+2}\right)d\omega.
\end{split}
\end{eqnarray}
For small $\alpha$, due to weak SOC, $g_y^+$ in Eq. (\ref{eq16}) can be neglected, only $g_n^+$ and $g_0^+$ survive. Regarding their contributions to the anisotropic spin model, $g_0^+$ simply acts as a background. Thus, the mechanism $\safebigding{172}$ works, i.e., $g_n^+$, contributed by altermagnetism, determines the resulting XXZ spin model ($J_{xx}=J_{yy}\neq J_{zz}$). As $\alpha$ increases, the effect of SOC becomes progressively more pronounced, leading to a non-negligible $g_y^+$. Note that the spin model solely contributed by $g_y^+$ is XYX-type, different from the original altermagnetism-induced XXZ spin model. Consequently, the mixture of XXZ and XYX results in an XYZ spin model. Physically, the effects of altermagnetism and SOC here are competitive, i.e., the mechanism $\safebigding{175}$ works.
\par
(2) $\alpha$-independent XYZ spin model (blue double underline in Table \ref{table3}). For small $\alpha$, similar to case (1), altermagnetism dominates the spin model (mechanism \safebigding[1.2]{172}). The critical distinction is that the spin model here is XYZ-type, which is attributed to the non-axial N\'{e}el vector orientation $\theta_n=\pi/3$, as indicated by Eq. (\ref{eq12}). Remarkably, this XYZ spin model persists even at large $\alpha$ because it supports the strongest anisotropy (i.e., $J_{xx}\neq J_{yy}\neq J_{zz}$). Such anisotropy can no longer be enhanced by SOC, since SOC merely generates an XYX spin model with moderate anisotropy (Table \ref{table1}) for impurities aligned along $\theta_R=0$;
\par
(3) $\alpha$-independent XYY spin model (red double underline in Table \ref{table3}). For $\theta_n=\pi/2$, the spin model solely contributed by altermagnetism is XYY-type [Eq. (\ref{eq12})]. The same spin model can also be induced by SOC if impurities are aligned along $y$-axis ($\theta_R=\pi/2$). In other words, by selecting proper N\'{e}el vector orientation and impurity configuration, altermagnetism and SOC coincidentally produce the same spin model (mechanism \safebigding[1.2]{176}). This coincidence maintains the spin model unchanged as $\alpha$ varies. Unlike case (2), the anisotropy of the spin model here is moderate.
\par
In brief summary, the spin model dominated by altermagnetism is primarily determined by the N\'{e}el vector orientation, while that dominated by SOC is dependent on the specific impurity configuration. Altermagnetism and SOC exhibit not only competition but also coincidence. Competition yields an XYZ spin model with maximal anisotropy, whereas coincidence produces a spin model with moderate anisotropy.
\par
For impurities aligned along $\theta_R=\pi/4$, the spin model at small $\alpha$ is always XXX-type, irrelevant to the N\'{e}el vector orientation. This is attributed to the vanishing altermagnetism at $\theta=\pi/4$ (as stated in Sec. III-A-1), which implies that only the kinetic term contributes to the RKKY components $J_{ii}$ (mechanism $\safebigding{174}$). At large $\alpha$, the results are sensitive to the N\'{e}el vector orientation, i.e., XXZ spin model for $\theta_n=0$ while XYZ model is induced once the N\'{e}el vector deviates from the $z$-axis. To understand this phenomenon, one has to examine the RKKY components $J_{ii}$ in Eq. (\ref{eq9}). Under the condition $\theta_R=\pi/4$, $J_{ii}$ can be approximated as
\begin{eqnarray}\label{eq17}
\begin{split}
J_{xx}&=-\frac{2J^{2}}{\pi }{\rm{Im }}\int\nolimits_{-\infty }^{u_{F}}\left[
g_{0}^{+}g_{0}^{-}+n_{x}\left(
g_{n}^{+}g_{x}^{-}+g_{n}^{-}g_{x}^{+}\right) \right] d\omega ,\\
J_{yy}&=-\frac{2J^{2}}{\pi }{\rm{Im }}\int\nolimits_{-\infty }^{u_{F}}\left[
g_{0}^{+}g_{0}^{-}-n_{x}\left(
g_{n}^{+}g_{x}^{-}+g_{n}^{-}g_{x}^{+}\right) \right] d\omega ,\\
J_{zz}&=-\frac{2J^{2}}{\pi }{\rm{Im }}\int\nolimits_{-\infty }^{u_{F}}\left[
g_{0}^{+}g_{0}^{-}-2g_{y}^{+}g_{y}^{-}-n_{x}\left(
g_{n}^{+}g_{x}^{-}+g_{n}^{-}g_{x}^{+}\right) \right]d\omega .
\end{split}
\end{eqnarray}
For $\theta_n=0$ ($n_x=0$), the last term in the square bracket of the above equation vanishes. Consequently, the spin model is determined by the SOC-induced term $-2g_{y}^{+}g_{y}^{-}$, which contributes to an XXZ spin model ($J_{xx}=J_{yy}\neq J_{zz}$), reminiscent of the case in the 2D electron gas \cite{electrongas}. As the N\'{e}el vector $\mathbf{n}$ deviates from the $z$-axis, a new term $n_{x}\left(
g_{n}^{+}g_{x}^{-}+g_{n}^{-}g_{x}^{+}\right)$, behaves as a XYY spin model, is generated. Naturally, a more complex spin model (XYZ) emerges due to the mixture of XXZ and XYY. The underlying physics can still be attributed to the mechanism $\safebigding{175}$, i.e., altermagnetism (realizing the XYY model via the in-plane component of $\mathbf{n}$) and SOC (inducing the XXZ model due to the configuration $\theta_R=\pi/4$) competitively shape the spin model.
\par
\begin{figure}[!htb]
\centering \includegraphics[width=0.35\textwidth]{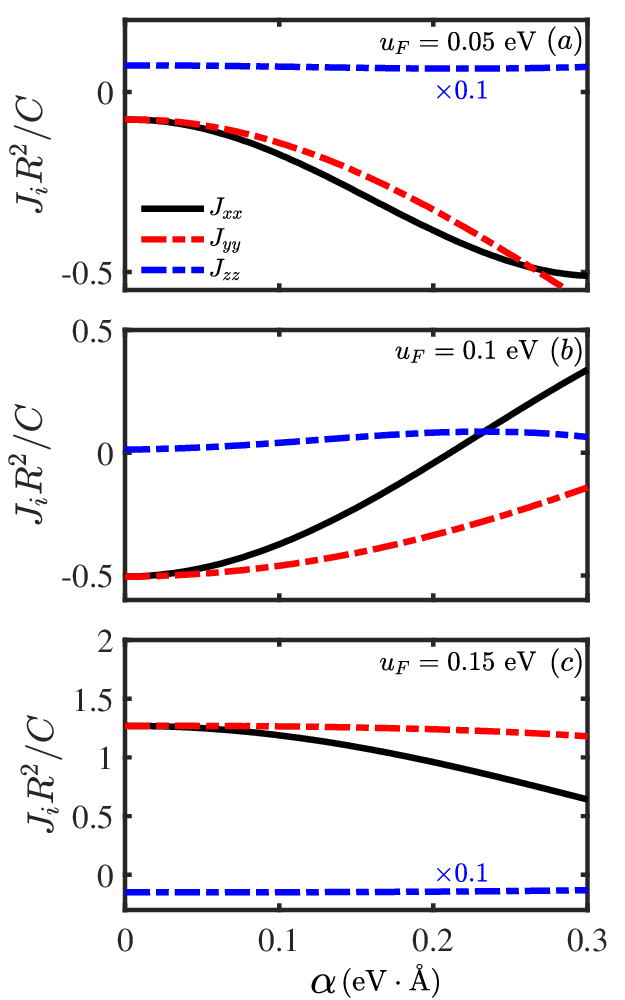}
\caption{(Color online) The RKKY components $J_{ii}$ ($i=x,y,z$) versus the SOC strength $\alpha$ for different Fermi energies: (a) $u_F=0.05$ ${\rm eV}$; (b) $u_F=0.1$ ${\rm eV}$; (c) $u_F=0.15$ ${\rm eV}$. The N\'{e}el vector is along the $z$-axis ( $\theta_n=0$, $\varphi_n=0$). Impurities are aligned along $\theta_R=0$ with spacing $R=30$ ${\rm \AA }$.}
\end{figure}
Note that the evolution of the spin model listed in Table \ref{table3} occurs even for the SOC strength $\alpha\in$ [0, 0.2 ${\rm eV}$ $\AA$], as shown in Figs. 2-4. Within this range of $\alpha$, Ref. \cite{alterRKKY3} only considers the effect of the zero-order term of $\alpha$ on the RKKY collinear components $J_{ii}$, i.e., completely neglecting the effect of SOC on the spin model. In fact, our research shows that when $\alpha>0.1$ ${\rm eV}$ $\AA$, the higher-order terms of $\alpha$ cannot be ignored, otherwise, as demonstrated in Refs. \cite{alterRKKY2, alterRKKY3}, the changes in the spin model would be masked. In addition, the regularity of the spin model in Table \ref{table3} is independent of the impurity distances (Figs. 2-4) and the Fermi energy $u_F$ (Fig. 5). This independence facilitates detection by eliminating the need for precise impurity spacing control or for specific selection of $u_F$.
\par
Taken together, our results demonstrate that the RKKY interaction serves as a sensitive magnetic probe for the intrinsic physics of pristine altermagnets. The directional lock between the Ising term and the N\'{e}el vector (in the absence of SOC), and the signatures of the N\'{e}el vector orientation in the DM components (in the presence of SOC), provide concrete methods to determine this crucial degree of freedom---the N\'{e}el vector orientation---magnetically. Furthermore, the emergence of the novel out-of-plane DM component $J_{DM}^z$ directly probes the fundamental synergy between two key ingredients: non-relativistic altermagnetism and relativistic SOC. The evolution of the RKKY spin model with SOC strength, governed by the five distinct mechanisms, essentially maps out how these two key ingredients compete or cooperate in shaping the magnetic anisotropy. Therefore, beyond describing impurity interactions in a "dirty" system, our study establishes the RKKY interaction as a powerful tool to uncover the physics of the host "perfect (pure)" material itself---specifically, its N\'{e}el vector orientation and the interplay between altermagnetism and SOC.

\section{Summary}
We have explored the effects of the N\'{e}el vector and Rashba SOC on the RKKY interaction in two-dimensional $d$-wave altermagnets. In the absence of SOC, we find a directional consistency between the N\'{e}el vector and the Ising term. This consistency enables direct determination of the N\'{e}el vector orientation $\mathbf{n}$ through detection of the Ising term. When SOC disrupts this consistency, the DM terms provide an alternative tool for qualitatively estimating the orientation $\mathbf{n}$. Intriguingly, a novel DM component, distinct from previous reports, emerges due to the synergy between altermagnetism and SOC. Additionally, we have studied how the RKKY spin model evolves with increasing SOC strength, revealing multifaceted effects on the spin model. Depending on the N\'{e}el vector orientation and the specific impurity configurations, altermagnetism may or may not contribute to the spin model. When not contributing, the spin model is determined by SOC or the kinetic term. When contributing, altermagnetism exhibits three distinct behaviors: it can solely dominate the spin model, coincidentally yield the same moderately anisotropic model as SOC, or compete with SOC to produce a spin model with maximal anisotropy. All results are numerically verified. These findings were inaccessible in previous literature, where studies considered only first-order SOC expansion with fixed N\'{e}el vector orientation. Our findings are particularly relevant for both predicted and realized altermagnetic systems, where the N\'{e}el vector can adopt either out-of-plane or in-plane orientations \cite{altermagnet3}. In such systems---especially when an altermagnet is grown on a substrate inducing Rashba SOC \cite{subs}, or when the SOC is tuned by a gate voltage \cite{alterRKKY3,gate}---phenomena such as the connection between the N\'{e}el vector and the Ising term without SOC, the emergence of novel DM terms, and the evolution of the spin model with SOC are expected to be observable. Our proposal is feasible with current techniques, such as spin-polarized scanning tunneling spectroscopy \cite{Laplane}, which can measure the magnetization curves of individual atoms, or electron spin resonance techniques coupled with optical detection schemes \cite{Wiebe1,Wiebe2}.

\section*{ACKNOWLEDGEMENTS} This work was supported by the National Natural Science Foundation of China (Grants No. 12104167, No. 12574050, No. 12274146, and No. 12274235), by the Guangdong NSF of China (Grant No. 2024A1515011300), by the Guangdong Basic and Applied Basic Research Foundation (Grant No. 2023B1515020050), and by the Guangdong Provincial Quantum Science Strategic Initiative (Grant No. GDZX2401002).

\section*{DATA AVAILABILITY} 
The data are not publicly available. The data are available from the authors upon reasonable request.

\end{document}